\documentclass{article}
\usepackage{arxiv}
\usepackage[utf8]{inputenc} 
\usepackage[T1]{fontenc}    
\usepackage{hyperref}       
\usepackage{url}            
\usepackage{booktabs}       
\usepackage{amsfonts}       
\usepackage{nicefrac}       
\usepackage{microtype}      
\usepackage{multicol}
\usepackage{graphicx}
\usepackage{adjustbox}
\usepackage{placeins}
\usepackage{algorithm} 
\graphicspath{ {./images/} }
\usepackage{caption}
\usepackage{subcaption}

\title{Human Emotion Classification based on EEG Signals Using Recurrent Neural Network And KNN}

\author{
 Shashank Joshi \\
 Department of Computer Science and Engineering\\
 SRM Institute Of Science And Technology\\
 Kattankulathur, Tamil Nadu – 603203, \\
 \texttt{sj8559@srmist.edu.in} \\
\And
 Falak Joshi \\
 SSJ Government Institute Of Medical Science And Research,\\
 Almora, Uttarakhand - 263620 \\
 \texttt{falakjoshi5756@gmail.com} \\
 }

\begin{document}
\maketitle
\begin{abstract}
\textbf{In human contact, emotion is very crucial. Attributes like words, voice intonation, facial expressions, and kinesics can all be used to portray one's feelings. However, brain-computer interface (BCI) devices have not yet reached the level required for emotion interpretation. With the rapid development of machine learning algorithms, dry electrode techniques, and different real-world applications of the brain-computer interface for normal individuals, emotion categorization from EEG data has recently gotten a lot of attention. Electroencephalogram (EEG) signals are a critical resource for these systems. The primary benefit of employing EEG signals is that they reflect true emotion and are easily resolved by computer systems. In this work, EEG signals associated with good, neutral, and negative emotions were identified using channel selection preprocessing. However, researchers had a limited grasp of the specifics of the link between various emotional states until now. To identify EEG signals, we used discrete wavelet transform and machine learning techniques such as recurrent neural network (RNN) and k-nearest neighbor (kNN) algorithm. Initially, the classifier methods were utilized for channel selection. As a result, final feature vectors were created by integrating the features of EEG segments from these channels. Using the RNN and kNN algorithms, the final feature vectors with connected positive, neutral, and negative emotions were categorized independently. The classification performance of both techniques is computed and compared. Using RNN and kNN, the average overall accuracies were 94.844 \% and 93.438 \%, respectively.
}
\end{abstract}


\section{Introduction}
Emotion is fundamental in human-human interaction and communication. 
Recognizing the emotional moods of others around us is a crucial aspect of natural communication. Given the prevalence of machines in our daily lives, emotional interaction between people and machines has emerged as one of the most pressing concerns in sophisticated human-machine interaction and brain-computer interface nowadays \cite{1}. While non-physiological attributes like words, voice intonation, facial expressions, and kinesics can all be used to portray one's emotion \cite{2}. In recent decades, many studies on emotion identification based on non-physiological attributes have been published \cite{3}.\\
Throughout the last few decades, numerous investigations on engineering techniques to automate emotion identification have been conducted. They can be divided into three categories. The first strategy focuses on facial expressions or speech analysis \cite{4}. These audio-visual approaches enable non-contact detection of emotion, causing no discomfort to the subject. However, these strategies may be more susceptible to deceit, and the settings might readily shift depending on the situation. The second method focuses on physiological signals at the periphery. Various studies suggest that alterations in the autonomic nervous system in the periphery, such as skin conductance (SC), electrocardiogram (ECG), respiration, and pulse, can be noticed in different emotional states \cite{5}. As compared to audio-visual-based techniques, peripheral physiological signal responses tend to provide more comprehensive and complicated information to evaluate emotional states.\\
The third strategy focuses on central nervous system brain signals such as electrocorticography (ECoG),  electroencephalography (EEG), and functional magnetic resonance imaging (fMRI). EEG signals, among various brain signals, have been shown to convey informative qualities in reactions to emotional states. EEG signals are obtained by monitoring voltage changes on the skull surface through electrical stimulation of activated neurons in the brain \cite{6}. In clinical settings, EEG is the most often utilized brain-activity-measuring technology for emotion recognition. Furthermore, EEG-based BCI devices open up a new channel of communication by detecting variations in the underlying pattern of brain activity while completing different tasks \cite{7}. However, BCI systems have yet to achieve the requisite level of emotion interpretation.\\
The interpretation of people's various emotional states via BCI systems, as well as automatic emotion recognition, may enable automated systems to react to humans in the future intuitively. They will be increasingly beneficial in a variety of disciplines, including medicine, entertainment, education, and many others \cite{8}. To understand emotions, BCI systems require variable resources that can be collected from humans and processed. Among the most essential resources for achieving this goal is the EEG signal. Emotion recognition is coupled with knowledge from several fields such as psychology, neurology, and engineering. SAM Self-Assessment Manikin (SAM)  surveys are commonly utilized in the creation of emotion detection systems for categorized affective responses of respondents \cite{9}. Furthermore, in this study, EEG patterns associated with positive, neutral and negative emotions were categorized.\\
The remainder of this paper is organized as follows: Related work is presented in Section 2. The employed methodology is depicted in Section 3. Section 4 describes the rationale behind our emotion induction experimental design. Section 5 presents a systematic description of methods like feature extraction, classification, etc. Section 6 discusses the experimental results and discussion. Finally, Section 7 concludes this paper.

\section{Related Work}
Several studies have employed various algorithms linked to EEG-based classification of emotional states. For example, Bhardwaj et al. \cite{10} highlights use of support vector machines (SVM) and LDA to classify seven emotions. In their investigation, three EEG channels (Fp1, P3, and O1) were utilized. The researchers looked into EEG signal subbands (theta, alpha, and beta). The overall average accuracies obtained were 74.13\% using SVM and 66.50\% using LDA.\\
Lee et al. \cite{11} distinguished between positive and negative emotions. In the proposed model the adaptive neuro-fuzzy inference system (ANFIS) yielded a classification accuracy of 78.45\%.\\
Channel et al. \cite{12} employed Fisher discriminant analysis (FDA) and Naive Bayes (NB) as the classification algorithms for the classification of positive and negative emotions. Classification accuracy was obtained as 70\% and 72\% for FDA and NB, respectively.\\
Davidson et al. \cite{13} present the use of a two dimensional model of emotion to categorize the emotions.\\
Murugappan et al. \cite{14} presents the classification of five emotions based on EEG signals. They captured EEG signals from different EEG channels and by employing techniques like kNN and linear discriminant analysis (LDA) algorithms, they attained maximum classification accuracy of 83.26\% and 75.21\%, respectively.\\
Zhang et al. \cite{15} employed Principal Component Analysis (PCA) for feature extraction.Two channels were used to extract the characteristics (F3 and F4). The researchers attained a classification accuracy of 73\%.\\

\section{Methodology}
The focal objective of this study was to categorize EEG signals associated with distinct emotions based on cinematic inputs using channel selection preprocessing. SAM was utilized to assess the emotional states of the individuals. Each audiovisual stimulus was judged by participants in terms of valence, arousal, like/dislike, dominance, and familiarity. EEG signals associated with good and negative emotions were categorized based on the valence assessments of the participants. The DWT approach was used to extract features from EEG recordings. EEG signal wavelet coefficients were believed to be feature vectors, and statistical characteristics were employed to minimize the dimension of those feature vectors. RNN and kNN algorithms were used to classify EEG data associated with positive, neutral, and negative emotion categories. The EEG channels with the best categorization performance were identified. As a result of merging the features of those EEG channels, final feature vectors were created. The classification of the final feature vectors was performed, and their performances were compared. Figure 1 depicts the processes used throughout the classification process.
\begin{figure}[h] 
    \centering
    \begin{center}
    \includegraphics[scale=0.4]{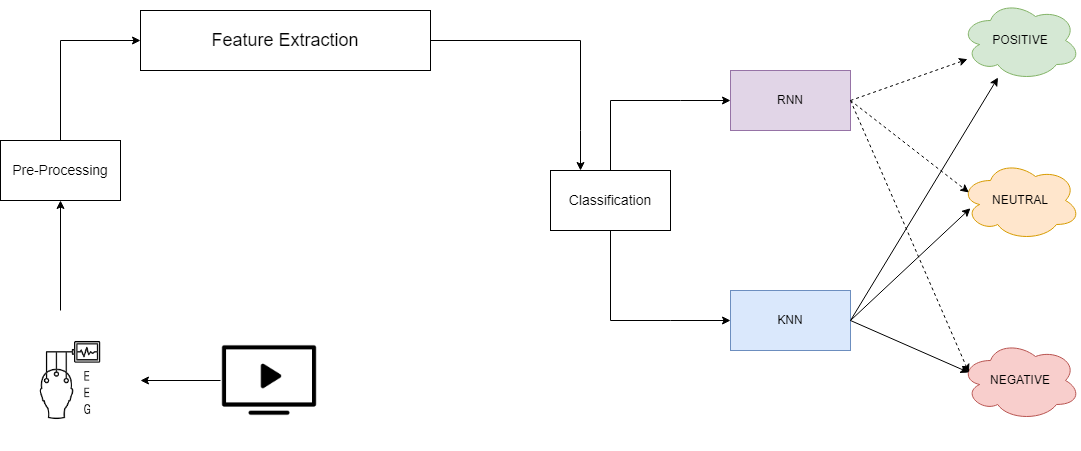}
    \end{center}
    \caption{Proposed Process of Classification}
\end{figure}

\FloatBarrier

\section{Materials}
\subsection{Database and Participants Questionnaire}
The publicly available database relating to emotional states was used in this investigation. Data was obtained from two participants (one male and one female) for three minutes per state (positive, neutral, and negative) \cite{15}. The  Muse EEG headband is used to record the EEG placements of the TP9, AF7, AF8, and TP10 through dry electrodes. The stimuli (Audio-Visual Clips) used to elicit the emotions are listed in table 1.
\FloatBarrier
\begin{table*}[h]
\begin{center}
\begin{adjustbox}{width=1\textwidth}
\setlength{\tabcolsep}{15pt}
\renewcommand{\arraystretch}{3}

\begin{tabular}{ |c|c|c| } 
\hline
\textbf{Clip No.} & \textbf{Audio-Visual Clip} & \textbf{Emotion} \\
\hline
1 & Slow Life - Nature timelapse & Positive \\ \hline
2 & Funny Dogs - Funny dog clips & Positive \\ \hline
3 & La La Land - Opening musical number & Positive \\ \hline
4 & Marley and Me - Death Scene & Negative \\ \hline
5 & Up - Opening Death Scene & Negative \\ \hline
6 & My Girl - Funeral Scene & Negative \\ \hline
\end{tabular}
\end{adjustbox}
\end{center}
\caption{Description of the Audio-Visual clips}
\label{tab:my_label}
\end{table*}
\FloatBarrier
\subsection{Task}
\begin{figure}[h] 
    \centering
    \begin{center}
    \includegraphics[scale=0.45]{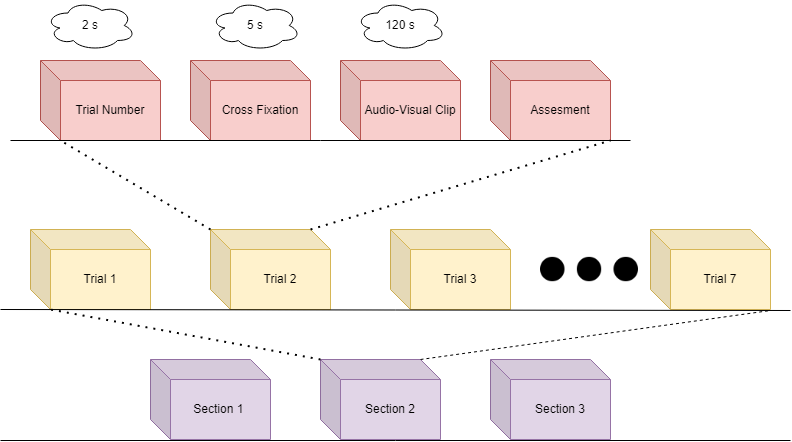}
    \end{center}
    \caption{Protocol to record EEG signals}
\end{figure}
Each participant's EEG signals were captured in a specific paradigm framework based on audiovisual stimuli. The paradigm is divided into two portions, each of which has seven trials, each of which includes the following steps:
\begin{enumerate}
    \item Trial Instance: A two-second screen that displays the current trial.
    \item Cross Fixation: A five-second fixation crosses.
    \item Audio-visual Clip: A 60-second exhibition of music snippets to elicit various emotions.
    \item Evaluation/ Participating Rating: SAM was used to assess valence, arousal, dominance, likes, and familiarity.
\end{enumerate}

Figure 2 depicts the paradigm used on participants to record EEG and peripheral physiological information. 
\subsection{EEG Signal Recording}
Electroencephalography is the process of obtaining electrophysiological data and signals from the brain using electrodes \cite{16}. Electrodes can be put subdural \cite{17}, that is, beneath the skull, on, and within the brain itself. Noninvasive approaches necessitate the placement of wet or dry electrodes around the cranium \cite{18}. Observing an instance at time t, raw electrical data is measured in microvolts ($\mu$V), resulting in wave patterns ranging from time period t to t+n. The EEG signal dataset is depicted by Figure 3.
\begin{figure}[h] 
    \centering
    \begin{center}
    \includegraphics[scale=0.7]{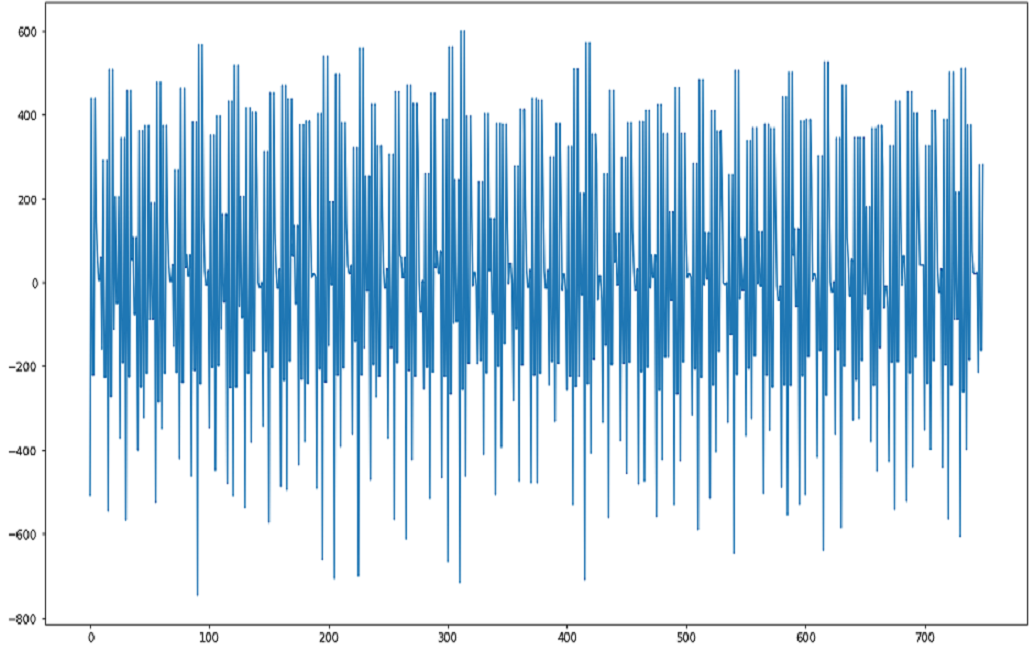}
    \end{center}
    \caption{EEG Signals Dataset (Amplitude vs Time)}
\end{figure}
\\
\\
\\
\section{Methods}
\subsection{Discrete wavelet transform (DWT)}
DWT is commonly utilized for non-stationary signal analysis \cite{6}. DWT has long been preferred for interpreting EEG signals. The DWT method can be used to get the appropriate frequency range. The DWT method divides the signal into sub-bands by filtering it with sequential high pass and low pass filters g[n] and h[n] in the time period \cite{19}. As a result, signal sub-bands with first-level d1 detail and a1 approximation are achieved. The definitions of detail sub-bands and first-level approximation  are as follows:
\begin{equation}
    d_1=\sum_{k=-\infty}^{\infty} x[k] . g[2n-k]
\end{equation}
\begin{equation}
    a_1=\sum_{k=-\infty}^{\infty} x[k] . h[2n-k]
\end{equation}
\subsection{Feature Extraction}
The DWT approach was used to generate feature vectors for EEG signals. 
The signal decomposition produces a collection of wavelet coefficients in each sub-band. The wavelet coefficients compactly reflect EEG features \cite{19}. Five statistical parameters were employed to reduce the dimensionality of the feature vector. The five statistical parameters i.e. a technique for reducing the dimension of feature vectors based on the theta band are as follows:
\begin{enumerate}
    \item The absolute maximum of the wavelet coefficients.
    \item The mean of the wavelet coefficients' absolute values.
    \item The coefficients' standard deviation.
    \item Coefficients' average power
    \item Coefficients' average energy
\end{enumerate}
Five-dimensional feature vectors corresponding to EEG segments connected to each emotional state were obtained at the end of the DWT approach, and statistical processes were utilized for feature extraction.

\subsection{Classification Algorithm}
RNN and kNN algorithms were used to classify the feature vectors of EEG data associated with positive and negative emotions. The performance of various pattern recognition algorithms was assessed and compared.
\subsubsection{k-nearest neighbor (kNN)}
The kNN algorithm is a distance-based classification technique. By studying the attributes of a new object, kNN can assign an entity encountered in (n) dimensional vector space to a specified class \cite{20}. The classification phase in the kNN algorithm begins with the calculation of the distance between a new object of an unknown class and each object in the training set. The classification process is completed by allocating the new object to the class that is most prevalent among its k-nearest neighbors. For classification, known classes of objects in the training set of k-nearest neighbors were used.\\
In the kNN algorithm, various methods are utilized to determine distance. 
Minkowski, Manhattan, Euclidean, and Hamming distance measures are examples of these methods. The Minkowski distance was utilized in this study to determine the nearest neighbors. The Minkowski distance is calculated as:
\begin{equation}
    distance = (\sum_{i=1}^{n}(x_i-y_i)^c)^\frac{1}{c}
\end{equation}
where x and y are the positions of points in a vector n-space.\\

Calculating the classification algorithm's ability with different k values optimizes the choice of k. To achieve the highest classification accuracy, the optimum value of k was discovered to be different for each participant.
\subsubsection{Recurrent neural networks (RNN)}
Recurrent neural networks (RNN) are a sort of NN that is extensively employed in the sequence analysis process because they are designed to extract contextual information by establishing the relationships between distinct time stamps. RNN is made up of many successive recurrent layers that are sequentially modeled in order to model the sequence with other sequences. RNN has an excellent aptitude for extracting contextual data from a sequence. The contextual cues in the network structure, on the other hand, are stable and successfully utilized to perform the data classification process. RNN may operate on sequences of any length. The architecture of the RNN classifier is depicted in Figure 3.
\begin{figure}[h] 
    \centering
    \begin{center}
    \includegraphics[scale=0.5]{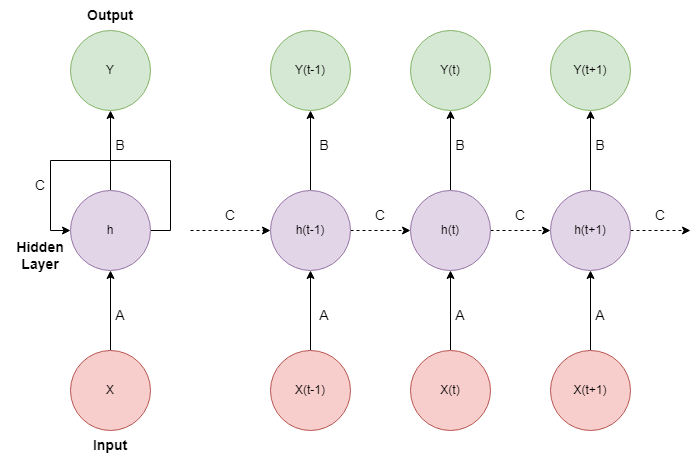}
    \end{center}
    \caption{Working of RNN (Recurrent Neural Network)}
\end{figure}
\subsubsection{Performance Evaluation}
The trained network's performance can be evaluated using accuracy, specificity, and sensitivity. These are some of the most commonly used success evaluation criteria in the literature. In this work, specificity indicates the capacity to correctly categorize samples including positive emotions, whereas sensitivity reflects the ability to accurately classify samples containing negative emotions. Eqs. (4), (5), and (6) calculate accuracy, specificity, and sensitivity, respectively.
\begin{equation}
    Accuracy = \frac{TN+TP}{TN+FN+TP+FP}
\end{equation}
\begin{equation}
    Specificity = \frac{TP}{TP+FP}
\end{equation}
\begin{equation}
    Sensitivity = \frac{TN}{TN+FN}
\end{equation}
Where,\\
TN: True Negative (Number of true decisions influenced by negative emotions)\\
TP: True Positive (Number of true decisions made by an automated system based on positive sentiment)\\
FN: False Negative (Number of false decisions influenced by negative emotions)\\
FP: False Positive (Number of false decisions made by an automated system based on positive sentiment)
\section{Experimental Results and Discussion}
To undertake a more efficient and reliable classification process, we created a training-set and a test-set for each of the model. The training set was composed of the first two sessions of the EEG dataset for each emotional state. The test set was composed of the most recent session of the EEG dataset for each emotional state. In Fig. 5(a),5(b),5(c), EEG signals related to positive,neutral and negative emotions state respectively are shown.
\begin{figure}[h]
     \centering
     \begin{subfigure}[b]{0.32\textwidth}
         \centering
         \includegraphics[width=\textwidth]{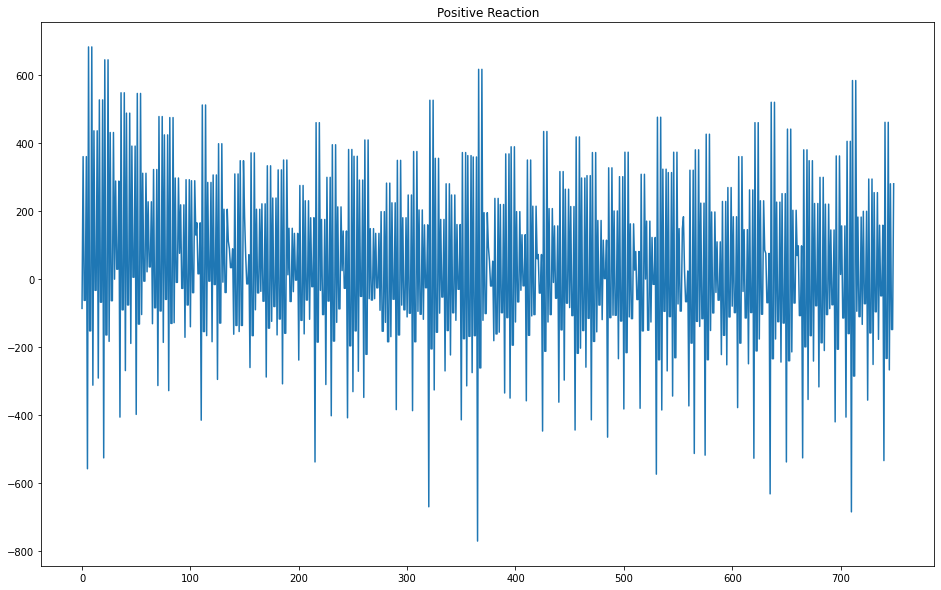}
         \caption{Positive Emotion}
         
     \end{subfigure}
     \hfill
     \begin{subfigure}[b]{0.32\textwidth}
         \centering
         \includegraphics[width=\textwidth]{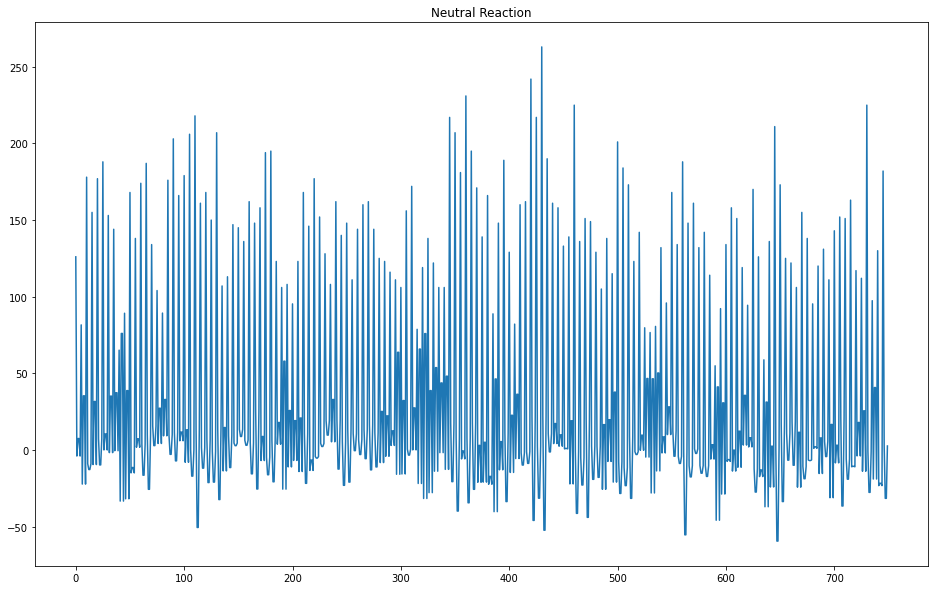}
         \caption{Neutral Emotion}
         
     \end{subfigure}
     \hfill
     \begin{subfigure}[b]{0.32\textwidth}
         \centering
         \includegraphics[width=\textwidth]{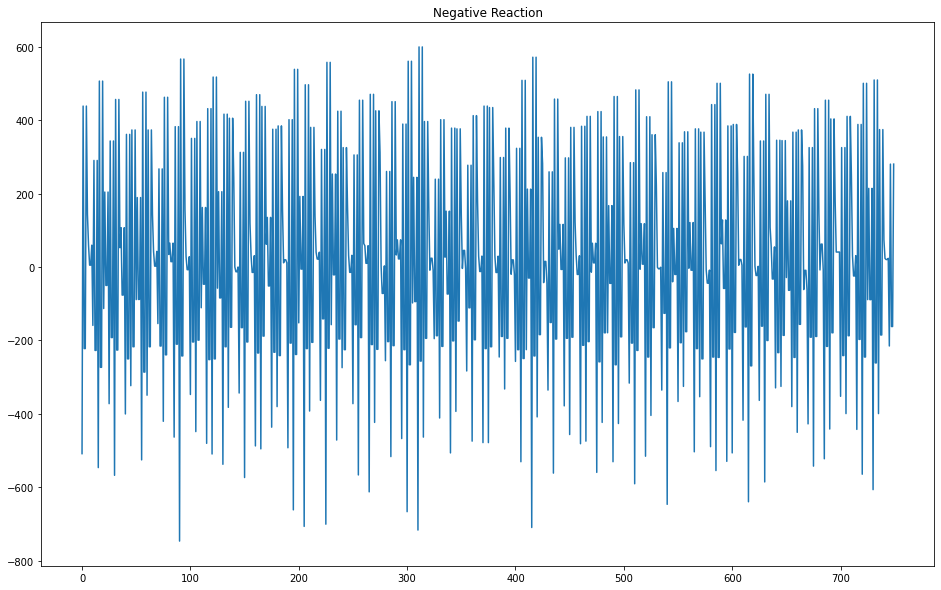}
         \caption{Negative Emotion}
         
     \end{subfigure}
        \caption{Categorized EEG Signals Dataset (Amplitude vs Time)}
        \label{fig:three graphs}
\end{figure}
\subsection{Classification of emotions by using RNN}
The final feature vectors were created by integrating the features of EEG segments from the chosen channels ( TP9, AF7, AF8, and TP10). As a result, for each EEG segment associated with positive, neutral, and negative emotions, new feature vectors made of 25 samples were produced. Furthermore, the classification process was applied to train the model, and the classification report and the confusion matrix of the model are shown in Figure 6.
\begin{figure}[H] 
    \centering
    \begin{center}
    \includegraphics{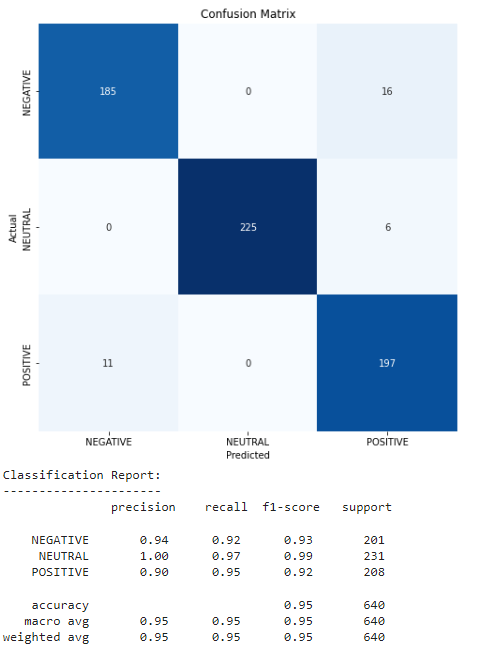}
    \end{center}
    \caption{Confusion Matrix And Classification Report Of RNN}
\end{figure}
\subsection{Classification of emotions by using kNN}
In this section, the kNN method was used to classify the final feature vectors acquired from EEG segments. The Minkowski distance measure was chosen to implement this classification. To improve the reliability of the classification results, the training and testing data were changed five times at random (fivefold cross-validation). Classification accuracies revealed that the k value yielding the lowest error value varied depending on the subject. From there, the k parameter is set to 1 or 3 for the classification process. Furthermore, the classification process was applied to train the model, and the classification report and the confusion matrix of the model are shown in Figure 7.
\begin{figure}[H] 
    \centering
    \begin{center}
    \includegraphics{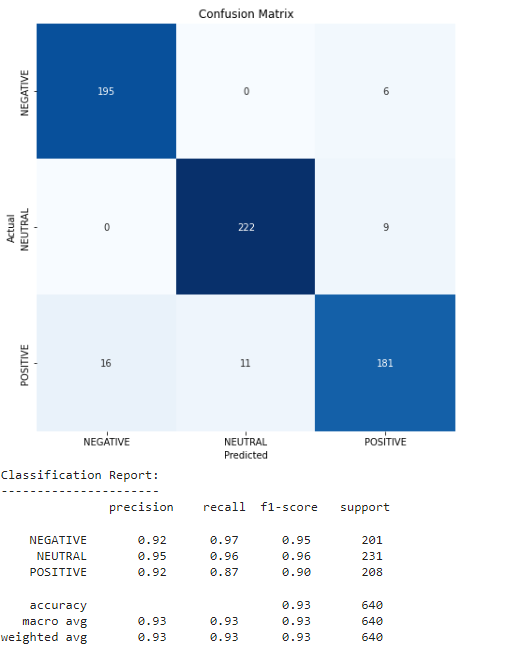}
    \end{center}
    \caption{Confusion Matrix And Classification Report Of kNN}
\end{figure}
This study's findings are discussed in more detail below:
\begin{itemize}
    \item The DWT (Discrete Wavelet Transform) method was used to obtain wavelet coefficients associated with various emotions (i.e. positive, neutral and negative). As feature vectors, these wavelet coefficients were examined. To reduce processing load, the size of feature vectors was lowered by applying five statistical parameters.
    \item RNN and kNN techniques were used to classify combined feature vectors from five channels ( TP9, AF7, AF8, and TP10). RNN was used in this study to determine the emotional state from EEG signals. In this work, the kNN technique was also utilized as a classifier to improve the dependability of the RNN results. kNN is one of the most basic and straightforward classification methods. The kNN algorithm is commonly utilized in numerous EEG applications. The fact that outcomes of the two algorithms were nearly identical. This matching contributes to the process's dependability.
    \item Statistical parameters (accuracy, specificity, and sensitivity) were averaged to determine the overall performance of RNN and kNN classifiers. Figure 8 depicts a comparison of averaged results for emotion classification. The efficiency of RNN was higher than that of kNN, as indicated in the figure. Both strategies, though, can be considered successful.
    \begin{figure}[H] 
    \centering
    \begin{center}
    \includegraphics[scale=0.4]{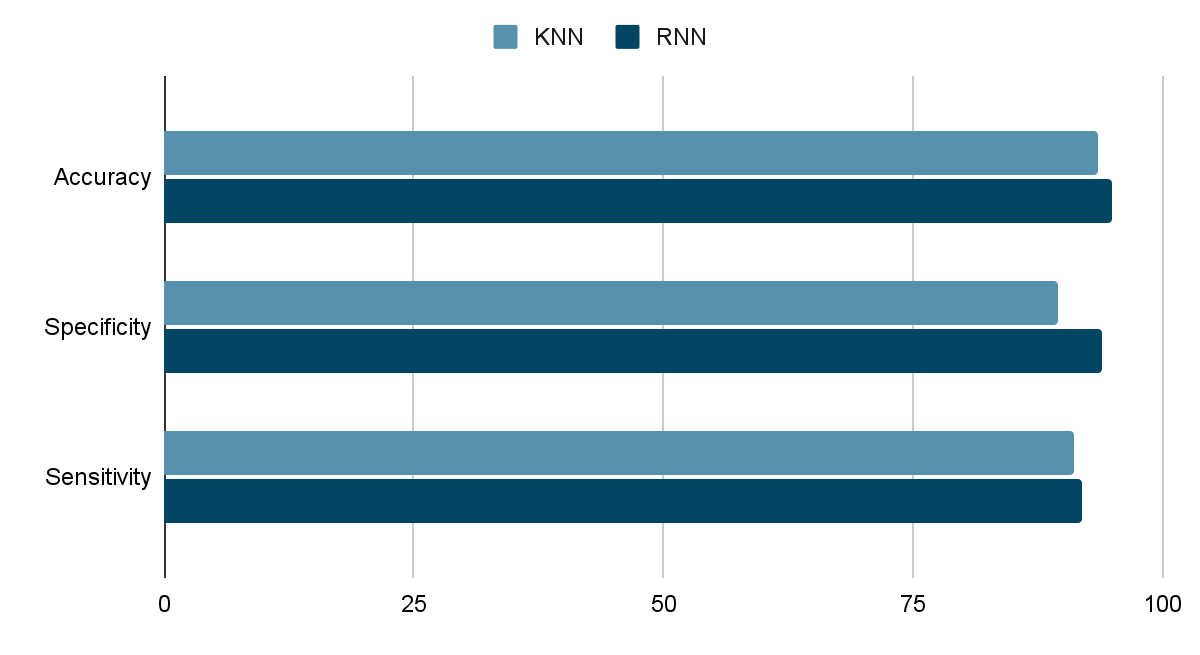}
    \end{center}
    \caption{Overall Performance Of RNN and KNN Classifiers}
\end{figure}
\item It was concluded that the RNN and kNN utilized in this investigation provide good accuracy results for emotion categorization.
\end{itemize}
\section{Conclusion}
In this study, we explored the characteristics of EEG data for emotion classification as well as a technique for tracking the trajectory of emotion changes. A series of tests employing movie clips is designed to elicit individuals' emotional responses, and an EEG data collection is obtained. For emotion classification, DWT was utilized for feature extraction, and RNN and kNN approaches were used as classifiers. The performance ranges of categorization for emotions are compatible with both RNN and kNN. As demonstrated by the findings, the critical points were a feature and channel decisions. It is believed that performance can be improved by using the proper channels and features.\\
The future development of this research will be focused on to improve classification efficiency, other physiological data such as blood pressure, respiration rate, body temperature, and GSR (galvanic skin reaction) can be combined with EEG signals.

\bibliographystyle{unsrt}  
\bibliography{references}  

\end{document}